\documentclass[aps,showpacs,amsmath,amssymbd,dvips,showkeys,preprint,preprintnumbers]{revtex4}

\usepackage{graphicx}
\usepackage{color}
\definecolor{DarkMagenta}{rgb}{0.55,0.00,0.55}

\def \be  {\begin{equation}}
\def \ee  {\end{equation}}
\def \bea {\begin{eqnarray}}
\def \eea {\end{eqnarray}}

\begin{document}

\preprint{ECTP-2015-03}
\preprint{WLCAPP-2015-03}
\vspace*{3mm}
\title{Chemical freeze-out in Hawking-Unruh radiation and quark-hadron transition}

\author{Abdel Nasser ~TAWFIK\footnote{http://atawfik.net/}}
\affiliation{Egyptian Center for Theoretical Physics (ECTP), Modern University for Technology and Information (MTI), 11571 Cairo, Egypt}
\affiliation{World Laboratory for Cosmology And Particle Physics (WLCAPP), Cairo, Egypt}

\author{Hayam Yassin}
\author{Eman R. Abo Elyazeed}
\affiliation{Ain Shams University, Faculty of Women for Arts, Science and Education, Physics Department, Cairo, Egypt}

\begin{abstract}

The proposed analogy between hadron production in high-energy collisions and Hawking-Unruh radiation process in the black holes shall be extended. This mechanism provides a theoretical basis for the freeze-out parameters, the temperature ($T$) and the baryon chemical potential ($\mu$), characterizing the final state of particle production. The results from charged black holes, in which the electric charge is related to $\mu$, are found comparable with the phenomenologically deduced parameters from the ratios of various particle species and the higher-order moments of net-proton multiplicity in thermal statistical models and Polyakov linear-sigma model. Furthermore, the resulting freeze-out condition $\langle E\rangle/\langle N\rangle\simeq 1~$GeV for average energy per particle is in good agreement with the hadronization process in the high-energy experiments. For the entropy density ($s$), the freeze-out condition $s/T^3\simeq7$ remains valid for $\mu\lesssim 0.3~$GeV. Then, due to the dependence of $T$ on $\mu$, the values of $s/T^3$ increase with increasing $\mu$. In accordance with this observation, we found that the entropy density remains constant with increasing $\mu$. Thus, we conclude that almost no information is going lost through Hawking-Unruh radiation from charged black holes. It is worthwhile to highlight that the freeze-out temperature from charged black holes is determined independent on both freeze-out conditions.

\end{abstract}


\maketitle
\makeatletter
\let\toc@pre\relax
\let\toc@post\relax
\makeatother

\section{Introduction}

It is widely accepted that the matter and radiation are confined inside black holes by gravitational force, i.e. they are enforced to remain within a restricted region of space \cite{0612151,Rees:2007,Karas:2007,0704.1426}. In such a gravitational confinement, isolation of the system is not literary absolute \cite{Hawking:1975}. In the strong fields at the outer edge of the black hole, the quantum excitation causes emission of Hawking-Unruh radiation \cite{Hawking:1974,Hawking:1975}. Such thermal emission leads to a slow decrease in the black hole mass and  eventually to an entire evaporation \cite{Hawking:1975}. The black hole mass decreases while the temperature increases. The latter can even reach the Planck scale. This makes the radiation process very similar to the Big Bang \cite{Carr:2005,Suzuki:2005,Harada:2006}.

Since no information transfer between inside and outside of the black hole is allowed, the Hawking-Unruh radiation must have {\it a priori} equal weights in all possible states of the black hole outside. It was concluded that the process at the stage of formation is thermal \cite{1401.0324}. From the pair production at the event horizon, the color neutrality (confinement) and the instability of the physical vacuum (to which the radiation is imported) allows transition through quantum tunneling and leads to the thermal radiation at the Hawking-Unruh temperature ($T_{HU}$) \cite{Ray:1992}. The latter can be determined in terms of the string tension ($\sigma$).

The quantum chromodynamics (QCD) possesses deconfinement property, as well \cite{Collins:1975}. Thus, it was suggested that the hadronization in high-energy collisions would be the analogue of the radiation from black holes  \cite{chapline,Recami:1976,severam,Grillo:1979}. Furthermore, the hadron production in high-energy experiments occurs through a succession of tunneling processes \cite{0704.1426}. In light of this, $T_{HU}$, which depends on the baryon number and angular momentum of the deconfined system \cite{0704.1426,0711.3712}, characterizes the QCD-analogue of the Hawking-Unruh radiation. The earlier could provide a dependence of $T_{HU}$ on the baryon chemical potential, while the angular momentum pattern of the radiation allows a centrality-dependence of $T_{HU}$ and eventually the elliptic flow \cite{0704.1426}. Femtoscopy and balance function are powerful tools for the temporal evolution of QCD hadronization \cite{TS2015}.

The nature of black hole beyond the Rindler horizon is given by the gauge theory gravity which is assumed to reproduce the dependence of the freeze-out temperature ($T_f$) on the center-of-mass energy ($\sqrt{s}$). String black hole was classified as a good candidate clarifying this dependence \cite{exact_string}. The qualitative dependence of  $T_f$ on $\sqrt{s}$ was definitively resolved \cite{Becattini:2006}. In some cases, Witten black hole explains fixity of $T_f$ in all elementary scattering processes at very large $\sqrt{s}$ \cite{exact_string}. Also, an effective description for the screening of the hadron string tension as function of $\mu$ and throughout an explanation for $T_f$ was proposed \cite{exact_string}.

Thus, Hawking-Unruh mechanism is conjectured to provide a theoretical basis for the production of newly formed hadrons in high-energy collisions and seems to be directly applicable at nearly vanishing baryon chemical potential \cite{1409.3104}. At $\mu=0$, it was found that this correspondence estimates the hadronization temperature $T_f=\sqrt{\sigma/2\, \pi} \simeq 165~$MeV  \cite{1409.3104}. Furthermore, the freeze-out conditions; $s/T_f^3=3\, \pi^2\, /4 \simeq 7.4$  and $\langle E \rangle / \langle N \rangle=\sqrt{2\, \pi\, \sigma}\simeq 1.09~$GeV agree well with the values deduced in the particle production in high-energy experiments \cite{1409.3104}.

The present paper is organized as follows. The correspondence of QCD hadronization and Hawking-Unruh radiation shall be reviewed in Sec. \ref{sec:corres}. The similarity between charged black hole and finite-density hadronization shall be investigated at finite baryon chemical potential in Sec. \ref{sec:chrgdBH}. Similar ideas have been proposed in Refs. \cite{0612151,0704.1426,0711.3712,exact_string}. The correspondence between black hole radiation and QCD hadronization temperature is determined in Sec. \ref{sec:tempr}. Thermodynamics of black hole and hadron shall be discussed in section \ref{sec:thrm}. In Sec. \ref{sec:frzout}, the freeze-out diagram and the two freeze-out conditions $s/T^3$ and $\langle E\rangle/\langle N\rangle$ shall be estimated and compared with the phenomenologically deduced parameters from the ratios of various particle species and the higher-order moments of net-proton multiplicity in thermal statistical models and Polyakov linear-sigma model (PLSM). Section \ref{sec:res} is devoted to the results and the discussion. The conclusions and outlook are elaborated in Sec. \ref{sec:cncl}.

\section{Correspondence of QCD hadronization and Hawking-Unruh radiation at vanishing and finite density}
\label{sec:corres}

A proposal that the gravitational confinement in primordial black holes might find analogue in the QCD confinement inside hadrons dates back to the seventies of the last century \cite{Recami:1976}. Due to color neutrality (confinement), the physical vacuum is thought to form an event horizon for quarks and gluons preventing even quantum tunneling \cite{0711.3712}. Since no information is transmitted through the black hole radiation to the outside, the radiation is likely thermal and must maintain color neutrality. A universal feature of $e^+\,e^-$, $p-\bar{p}$,  $p-p$ and  even heavy-ion collisions is that the particle production follows a thermal pattern and apparently occurs at the same temperature. The thermalization in heavy-ion collisions is obvious because of the possible rescatterings between the large number of interacting partons, which is not the case in $e^+\,e^-$ annihilation.

The hadronization is assumed as the QCD counterpart of Hawking radiation \cite{0704.1426}. This analogue solves many puzzles of thermalization, and universal freeze-out temperature and shall be utilized to explain various freeze-out conditions.
\begin{itemize}
\item Quantum tunneling through quark and gluon event horizon because of vacuum instability through pair production leads to thermal radiation at  a temperature determined by the string tension.
\item A succession of such tunneling processes is likely at high energy. Partition processes and resulting cascades lead to a limiting temperature as that in the thermal models.
\item In kinetic thermalization, the initial state information is likely lost through successive collisions, while stochastic QCD Hawking radiations are in equilibrium preventing information transfer.
\end{itemize}

For sake of completeness, it is worthwhile to add that even such proposed analogue might not be literary exact. This partly defines possible uncertainties in the present calculations. While the initial state information is likely lost through the strong collisions and kinetic thermalization \cite{0704.1426}, the stochastic QCD Hawking-Unruh radiation remains in equilibrium, because the quantum tunneling does not allow information transfer \cite{0704.1426}. 

As introduced in \cite{exact_string}, the correspondence between QCD hadronization and Hawking-Unruh radiation is a kind of phenomenological matching based on gauge theory gravity. The black hole behind which Rindler horizon could confirm the experimental dependence of $T$ on the center-of-mass energy ($\sqrt{s_{NN}}$), is likely a Witten black hole, especially the constant $T$ for all elementary scattering processes at large energies. Furthermore, an effective description of the screening of string tension ($\sigma$) with varying $\sqrt{s_{NN}}$ or chemical potential ($\mu$) was introduced \cite{exact_string}.

The extension to a finite chemical potential ($\mu$) should add a new {\it "charge"} introducing a modification in the quantum tunneling process and in the corresponding Hawking-Unruh hadronization temperature \cite{0711.3712}. The temperature of black hole radiation is determined by confinement phase transition. On the other hand, the dependence of the hadronization temperature on the baryon density forms a kind of {\it"white"} hole in high-energy collisions from which {\it "colored"} quarks are emitted. These are conjectured to remain confined by chromodynamic forces or equivalently by the bag pressure $B$. At nonvanishing baryon chemical potential ($\mu$) the Fermi repulsion between the corresponding quarks becomes very likely. This additional interaction appears in the form of density-depending pressure $P(\mu)$. The effects of baryon chemical potential are opposite to that of $B$.

\section{Rindler horizon and thermalization}

In the particle production, it is widely accepted that the nature of the freeze-out temperature is in statistical thermal equilibrium. On the other hand, the Rindler horizon represents another pillar for the proposed Hawking-Unruh radiation correspondence with the QCD hadronization. In $q\bar{q}$ pair production, the analogue of Rindler horizon is {\it "color-blind"} process which is produced by color-charge neutrality or confinement, i.e. Rindler-Unruh excitement of QCD vacuum. This means that the $q\bar{q}$ pair production are likely thermal  because of the random distribution of $q$ and $\bar{q}$, which are entangled in such QCD vacuum. The produced hadrons are assumed to be {\it "born in equilibrium"}.

The hadronic Rindler spacetime is considered as near-horizon approximation of specific black hole spacetimes. The existence of common thermodynamical features of both QCD hadronization and black hole radiation is proposed as a category defining that black hole spacetime that assures correspondence with particle the production \cite{exact_string}, especially the possibly to predict the dependence of $T$ on $\sqrt{s_{NN}}$ or $\mu$. Additionally, various conditions have been proposed in \cite{exact_string}; (i) proportionality of black hole mass to $\sqrt{s_{NN}}$, (ii) coincidence of string tension and coupling constant and (iii) correspondence of Hawking temperature with Unruh temperature. Furthermore, the authors of \cite{exact_string} added a fourth condition that the black hole partition function should diverge at a given temperature. According to this additional condition, they excluded most of the well-known black holes, such as Schwarzschild or Reissner-Nordstr\"om black hole. In proposing this condition, it was assumed  that the black hole radiation is a continuous process that lasts untill the black hole mass entirely evaporates. There is a great number of theoretical works examining the conundrum of the black hole evaporation, for instance \cite{giddings}. Indeed, the evaporation process that fulfills causality and thermodynamical consistency ends up with massive remnants. Accordingly, the partition function should be diverging, i.e. becoming singular, and the related Hawking temperature, which is inversely proportional to the remaining black hole mass, defines the {\it critical} temperature.

Unruh temperature is proportional to the acceleration ($a$) of a test particle in vicinity of the black hole horizon, which is assumed to emerge due to causally disconnected spacetimes, i.e. Rindler coordinate, $\eta=\ln(|(t+x)/(t-x)|)/2$, which is known as spacetime rapidity in the high-energy collisions \cite{tuchin}. For completeness' sake, the surface of Rindler space reads $\rho^2=x^2-t^2$.  In general relativity theory, Rindler space is considered as an approximation to the Schwarzschild metric of a large black hole. A rearrangement of the vacuum structure is possible through Bogoliubov transformation which relates operators for particle creation and annihilation in Rindler and Minkowski spaces. Furthermore, the vacuum in Minkowski space is related to that in the Rindler space by Bogoliubov transformation, in which the Rindler vacuum is populated with temperature $T=2\pi/a$. By considering chiral symmetry restoration induced by a rapid deceleration of the colliding nuclei, it was argued that the parton saturation in the initial nuclear wave functions is a necessary precondition for the formation of quark-gluon plasma (QGP), which can then hadronize \cite{tuchin}.

\section{Charged black holes and hadronization at finite density}
\label{sec:chrgdBH}

The black hole event horizon is created by the strong gravitational attraction, which leads to a diverging Schwarzschild metric at a certain value of the spatial extension ($R$) \cite{0704.1426}. At the equator, the invariant spacetime length element is given as
\be
ds^2 = \left(1 - 2\frac{GM}{R_S}\right)~\!dt^2 - \left(1- 2\frac{GM}{R_S}\right)^{-1}\; dr^2,
\label{Schwarz}
\ee
where $r$ and $t$ are flat space and time coordinates, respectively. At Schwarzschild radius $R_S = 2GM$, $d s$ diverges, i.e. become singular. In Eq. (\ref{Schwarz}), natural units are assumed ($c=\hbar=G=1$). To fit with the approach introduced in Ref.  \cite{1409.3104}, $G=1/(2\sigma)= 2.6316$, where $\sigma=0.19\;GeV^2$ defines the string tension. Thus, hereafter $G$ is not omitted from the expressions.

If the black hole has a net electric-charge $Q$, the resulting Coulomb repulsion weakens the gravitational attraction. This in turn modifies the event horizon \cite{Ray:1992} and  the corresponding line element (Reissner-Nordstr\"om metric) as well
\be
ds^2 = \left(1 - 2\frac{G M}{R_{RN}} + \frac{G Q^2}{R_{RN}^2}\right)\; d t^2 - \left(1- 2\frac{G M}{R_{RN}} + \frac{G Q^2}{R_{RN}^2}\right)^{-1}\; dr^2. 
\label{RN}
\ee
The field strength of the interaction is obtained from the coefficient
\bea
f(R_{RN}) &=& 1 - 2 \frac{G M}{R_{RN}} +  \frac{G Q^2}{R_{RN}^2}.
\eea
The divergence leads to  smaller Reissner-Nordstr\"om radius
\be
R_{RN} = R_S \left[\frac{1}{2}\left(1 + \sqrt{1 - \frac{Q^2}{G\, M^2}}\right)\right].
\label{RNradius}
\ee
At vanishing charge, $R_{RN}$ is reduced to the Schwarzschild radius ($R_S$).

\subsection{Radiation / hadronization temperature}
\label{sec:tempr}

The charged black hole radiation temperature is to be estimated from $f'(R_{RN})/4\pi$, where $f'(R_{RN})=\partial f(R_{RN}/\partial R_{RN})$ \cite{book_web},
\bea
T &=& \frac{1}{2 \pi} \left(\frac{G M}{R_{RN}^2} - \frac{G Q^2}{R_{RN}^3}\right). \label{eq:Tt}
\eea
Then, from Eqs. (\ref{RNradius}) and (\ref{eq:Tt}), the temperature of Hawking-Unruh radiation is given as
\bea
T_{BH}(M,Q) &=& \frac{G M R_{RN} - GQ^2}{2\pi R_{RN} G^2 M^2} \left(1+\sqrt{1-\frac{Q^2}{G\,M^2}}\right)^{-2} 
= T_{BH}(M,0)\; \left[\frac{4~\left(1 - \frac{Q^2}{G\, M^2}\right)^{1/2}}{\left(1 + \sqrt{1- \frac{Q^2}{G\, M^2}}\right)^{2}}\right],
\label{T-Q}
\eea
where $T_{BH}(M,0)$ is the Hawking-Unruh radiation temperature from Schwarzschild black hole, i.e. nonrotating and uncharged black hole. Similar expression has been obtained in Refs. \cite{Ruf,I-U-W}. It has been shown that the value of $T_{BH}(M,0)$ is in good agreement with the QCD freeze-out temperature, $T_{BH}(M,0) \simeq 175\pm15~$ MeV \cite{1409.3104}. It is obvious that $Q=0$ results in $T_{BH}(M,0)$ and increasing $Q$ [alternatively $\mu$, Eq. (\ref{muQ})] allows the quantities in the bracket of Eq. (\ref{T-Q}) to generate a parabola, Fig. \ref{Fig1}.

\subsection{Black hole / hadron thermodynamics}
\label{sec:thrm}

From the black hole mechanics, which has the same structure as the thermodynamics \cite{Bardeen:1973},
\begin{itemize}
\item due to Hawking-Unruh radiation, the black hole mass modification is given as
\be
d M = T\; d S + \mu\; d Q, \label{eq:1stThrm}
\ee
where $S$ is the entropy. It is apparent that Eq. (\ref{eq:1stThrm}) is nothing but the first law of thermodynamics in a grand canonical ensemble with finite $\mu$ and finite charge number $d Q$ \cite{0007195}. The parameter associating with a variation of the charge $Q$ is the baryon chemical potential $\mu$ \cite{0007195},
 \bea
 \mu 
  &\simeq & 2\, \pi\, \frac{Q\, R_{RN}}{G\, S} = 2\frac{Q}{R_{RN}}.
 \label{muQ}
 \eea
 Almost, the same expression (except of an additional factor $2$) was proposed in Ref. \cite{0704.1426}.
\item Also, from the second law of black hole mechanics,  the analogy of area and entropy, the Bekenstein-Hawking area law  reads \cite{Hawking:1971,Bekenstein}
\be
S = \pi \frac{R^2}{G}.
\label{entropy}
\ee
In a straightforward matter $R$ can be replaced by $R_{RN}$. Then Eq. (\ref{muQ}) can be modified to an entropy associating with the charged black hole radiation or equivalently with the hadron production at finite baryon chemical potential. The entropy density $s=S/V$. Assuming that the black hole is shaped as sphere, the volume is simply given as $V=4\, \pi\, R^3/3$.

\end{itemize}

\subsection{Freeze-out conditions}
\label{sec:frzout}

The proposed freeze-out temperatures are calculated from charged black holes. Of course, they are estimated independently on the freeze-out conditions. For quark-hadron phase-transition, there are various conditions describing the freeze-out diagram \cite{jeanRedlich,nb01,percl,Tawfik:2005qn,Tawfik:2004ss,Tawfik:2013dba,Tawfik:2013eua,fohm}. All of them are based on phenomenological descriptions, where the thermal statistical models are implemented in deriving the freeze-out parameters ($T$ and $\mu$). In doing this, the experimentally measured ratios of various particle species \cite{jeanRedlich,nb01,Tawfik:2005qn,Tawfik:2004ss,Tawfik:2013dba,Tawfik:2013eua} and the higher-order moments of net-proton multiplicity \cite{fohm} should be reproduced from the thermal statistical models through varying $T$ and $\mu$. These two parameters, at which the measured ratios agree well with the calculated ones, characterize the chemical freeze-out parameters corresponding to the concrete interacting system, at the given centrality, rapidity, and energy, etc.

\begin{itemize}
\item In the particle production at vanishing baryon chemical potential, the entropy density normalized to $T^3$ is proposed to have a constant value \cite{Tawfik:2005qn,Tawfik:2004ss}. Also, in the Hawking-Unruh radiation from uncharged black holes \cite{1409.3104}, we find that
\bea
\left.\frac{s}{T^3}\right|_{Q\; \text{or}\; \mu =0} = \frac{3}{8\, G^2\, M\, T^3_{BH}(M,0)}. \label{eq:sT3bh}
\eea
It was proved that the value of $s/T^3$ calculated from Schwarzschild black hole, Eq. (\ref{eq:sT3bh}), is almost identical to that from the particle production \cite{1409.3104}.

In quark-hadron phase transition, $s/T^3$ is assumed to remain constant with increasing $\mu$ \cite{Tawfik:2005qn,Tawfik:2004ss,Tawfik:2014eba}. In the present work, we show that the proposed correspondence of QCD hadronization and Hawking-Unruh radiation is valid at finite density. In other words, constant $s/T^3$ that was found in QCD hadronization has a correspondence in Hawking-Unruh radiation at finite $\mu$ or finite $Q$
 \bea
\left.\frac{s}{T^3}\right|_{Q\; \text{or}\; \mu \neq 0} = \left.\frac{s}{T^3}\right|_{Q\; \text{or}\; \mu =0} \left[\frac{2 \left(1+\sqrt{1-\frac{Q^2}{G\, M^2}}\right)^5}{\left(1-\frac{Q^2}{G\, M^2}\right)^{3/2}}\right]. \label{eq:sT3b}
 \eea
It is obvious that small $Q$ makes the value of the bracket in Eq. (\ref{eq:sT3b}) nearly unity. But, at large $Q$, the nominator increases faster than the denominator. More details shall be elaborated in the section that follows.

\item As proposed in \cite{1409.3104}, the average energy per particle at vanishing $\mu$ reads
\bea
\left.\frac{\langle E\rangle}{\langle N\rangle}\right|_{\mu=0}  &=& \sigma\, R_s. 
\eea
At finite $\mu$ or finite $Q$, it is assumed that that the string tension of charged black holes show the same phenomenological behavior \cite{exact_string}
\bea
\sigma(\mu) &\simeq & \sigma(\mu=0)\left[1-\frac{\mu}{\mu_0}\right],
\eea
where $\mu_0\simeq 1.2~$GeV. Also, at finite $\mu$  or finite $Q$, $R_s$ should be replaced by $R_{RN}$, Eq. (\ref{RNradius}). Then, the freeze-out condition becomes
\bea
\left.\frac{\langle E\rangle}{\langle N\rangle}\right|_{\mu\neq0} &=& \sigma(\mu)\, R_{RN}.
\eea
\end{itemize}

\section{Results and discussion}
\label{sec:res}

\begin{figure}[hbt]
\includegraphics[width=8.cm,angle=-90]{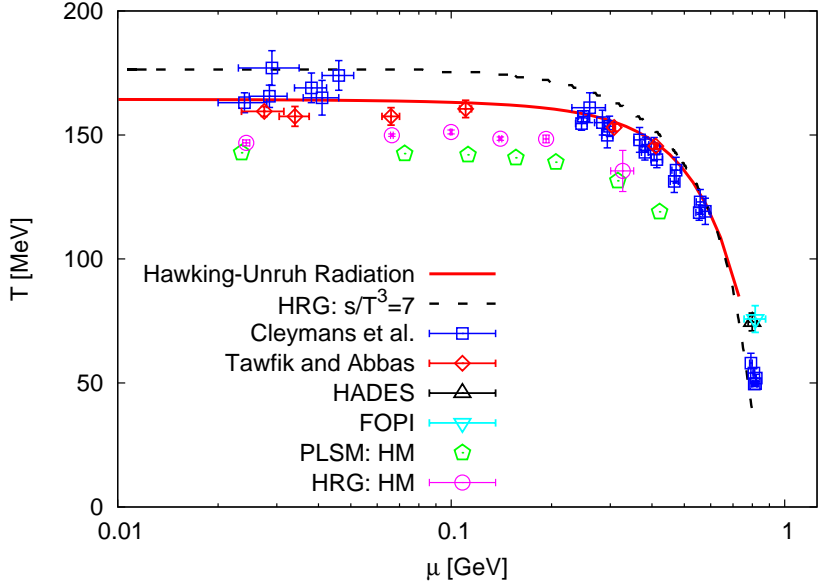}
\caption{The analogy of Hawking-Unruh radiation to the particle production (hadronization) is utilized in deducing the freeze-out parameters, temperature $T$, and baryon chemical potential $\mu_b$ (solid curve). The symbols refer to the experimentally deduced freeze-out parameters from particle ratios: Cleymans {\it et al.} \cite{clmns}, Tawfik and Abbas \cite{Tawfik:2013bza,Tawfik:2014dha}, HADES \cite{hds}, and FOPI \cite{fopi} and higher-order moments:  SU(3) Polyakov linear-$\sigma$ model (PLSM) \cite{AM} and HRG \cite{fohm}. \label{Fig1} }
\end{figure}

In Fig. \ref{Fig1}, the freeze-out parameters $T$ and $\mu$ (solid line) calculated from Eq. (\ref{T-Q}), where the equivalence of $Q$ and $\mu$ is given in Eq. (\ref{muQ}), are compared with the results from the hadron resonance gas (HRG) model at $s/T^3=7$ (dashed line) and the phenomenologically deduced parameters (symbols) from the particle ratios: Cleymans {\it et al.} \cite{clmns}, Tawfik and Abbas \cite{Tawfik:2013bza,Tawfik:2014dha}, HADES \cite{hds}, and FOPI \cite{fopi} and the higher-order moments:  SU(3) Polyakov linear-sigma model (PLSM) \cite{AM} and HRG \cite{fohm}. Despite the relative low freeze-out temperature from the higher-order moments, both qualitative and quantitative comparisons are obviously convincing. It is worthwhile to notice the excellent agreement with the recently estimated freeze-out parameters based on the RHIC beam energy scan program of the STAR experiment \cite{Tawfik:2013bza,Tawfik:2014dha}. Because of the anomaly in the proton-pion ratio at the large hadron collider (LHC) energies, we excluded their freeze-out parameters. With anomaly one refers to the particle ratio overestimation relative to the thermal statistical models at LHC energies. But, it is very likely that the freeze-out temperatures at very high energies (very low baryon chemical potential) likely remain almost unchanged as was measured at the relativistic heavy-ion collider (RHIC) top energies.

\begin{figure}[hbt]
\includegraphics[width=8.cm,angle=-90]{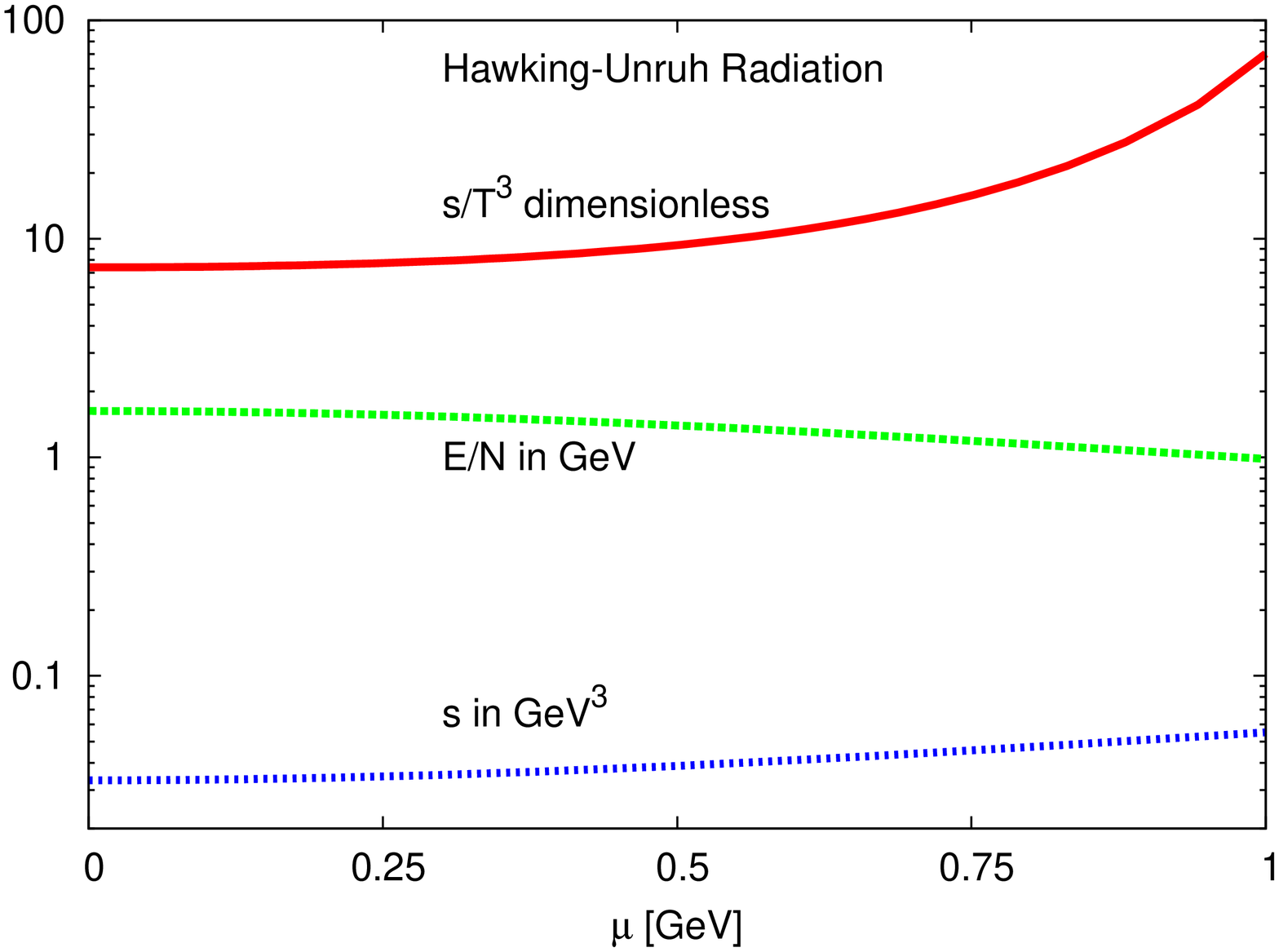}
\caption{By using the analogy of Hawking-Unruh radiation from charged black holes and the particle production in high-energy collisions, the freeze-out parameters $s/T^3$ (solid curve) and $\langle E\rangle/\langle N\rangle$ (dashed curve) and the entropy density $s$ (dotted curve) are given as a function of the baryon chemical potential, $\mu$. \label{Fig2}  }
\end{figure}

In a log scale, Fig. \ref{Fig2} gives the freeze-out conditions $s/T^3$ (solid curve) and $\langle E\rangle/\langle N\rangle$ (dashed curve) and the entropy density ($s$) (dotted curve) as a function of  the black hole charges or equivalently the baryon chemical potentials. It is obvious that the entropy density normalized to $T^3$ remains constant ($\sim 7$) at $\mu\lesssim 0.3~$GeV. It is worthwhile to highlight that this is the same range, in which the freeze-out temperature is almost independent on $\mu$, Fig. \ref{Fig1}. This observation agrees well with the estimation deduced from the ratios of hadron species in high-energy collisions \cite{Tawfik:2005qn,Tawfik:2004ss}.

The values of $s/T^3$ deduced from Eq. (\ref{eq:sT3b}) increase with increasing $\mu$ (or $Q$). This can be understood from the expression  (\ref{eq:sT3b}). Relative to the denominator, the numerator has a higher exponent. This makes the fraction in the bracket considerably increases with $Q$ (or $\mu$). When comparing $s/T^3$ from the black hole radiation with the QCD freeze-out temperature, Fig. \ref{Fig1}, and with $\left.T_{BH}\right|_{\mu \neq 0}$, Eq. (\ref{T-Q}), the uncertainties in $\left.s/T^3\right|_{\mu \neq 0}$, Eq. (\ref{eq:sT3b}), seems to play an essential role in giving a partial explanation why $\left.T_{BH}\right|_{\mu \neq 0}$ perfectly agrees with the results from other models and various heavy-ion experiments while $\left.s/T^3\right|_{\mu \neq 0}\simeq 7$ varies with $\mu$ (or $Q$). Other effects shall be elaborated on shortly.

The Hawking-Unruh temperature ($T$) describes well the quark-hadron freeze-out results, Fig. \ref{Fig1}. The proposed value of $s/T^3$ is well found at black hole mass $M=1.629~$GeV. The radiation temperature is inversely proportional to $M$ \cite{Hawking:1975}. Therefore, two possibilities can be proposed.
\begin{enumerate}
 \item The entropy density ($s$) would be responsible for the increase in $s/T^3$ with $\mu$ (or $Q$). This would be interpreted as the extensive quantity $s$ carries various information as well as thermodynamic properties of the system of interest.
 \item Alternatively, $s$ would remain constant with $\mu$ (or $Q$), Fig. \ref{Fig2}, while $T$  depends on $\mu$ (or $Q$) given in Eq. (\ref{T-Q}) and illustrated in Fig. \ref{Fig1}.
 \end{enumerate}
The intensive quantity $T$ is uniquely defined and was determined not depending on the black hole radiation freeze-out conditions, $s/T^3$ and $\langle E\rangle/\langle N\rangle$. So far, we conclude that while $T$ seems to be accurately estimated, Fig. \ref{Fig1}, $\left.s/T^3\right|_{\mu \neq 0}$ does not, Fig. \ref{Fig2}. The second possibility is verified in Fig. \ref{Fig2}. The entropy density ($s$) is indeed almost independent on $\mu$ (or $Q$). Thus, we conclude that the Hawking-Unruh radiation from charged black holes is not associated with information loss. Also, we conclude that the degrees of freedom seem to remain conserved throughout the radiation process, while the charges of black holes do not matter. 

Furthermore, it is apparent that the average energy per particle $\langle E\rangle/\langle N\rangle$ remains nearly constant over the whole range of $\mu$ (or $Q$), $\simeq 1.1~$GeV. This result agrees well with the value proposed in Ref. \cite{jeanRedlich}.

\section{Conclusions and outlook}
\label{sec:cncl}

Assuming that in the early Universe, the fluctuations in the space-time metric are scale invariant, the speed of sound in the primordial fireball must exceed $1\%$ the speed of light, as a consequence of observational limits on primordial black holes. The speculations about the black hole formation in the LHC attracts the attention of many theoreticians \cite{bhLHC1,bhLHC2}.

In the present work and in connection with black holes with gauge charges and their thermodynamics, the analogies between the black hole  properties and those of hadrons are investigated. Concretely, we present an extension to a recent proposal that for high-energy hadron productions, the freeze-out parameters can be interpreted as Hawking-Unruh radiation. We propose the existence of a universal hadronic freeze-out diagram based on the Hawking-Unruh radiation temperature.

Assuming that the black hole charge can be related to the baryon chemical potential, we have compared the resulting temperature with the freeze-out parameters deduced from the ratios of various particle species and the higher-order moments of net-proton multiplicity measured in different high-energy experiments and confronted to the hadron resonance gas and Polyakov linear-sigma models. We found that the agreement is very convincing. This is an evidence that the freeze-out parameters, $T$ and $\mu$, can be interpreted by the gravitational deconfinement and the Hawking-Unruh radiation from charged black holes.

We have estimated two freeze-out conditions, $s/T^3$ for entropy density $s$ and  $\langle E\rangle/\langle N\rangle$ for average energy per particle in charged black holes. The latter is directly expressed in terms of the finite-density string-tension and the spacial extension of the charged black hole, while the earlier is calculated in dependence of $s/T^3$-value calculated from Schwarzschild black hole multiplied by a function of charge and radius of the charged black hole. It is found that $\langle E\rangle/\langle N\rangle$ remains almost constant over the whole range of $\mu$, while $s/T^3\simeq 7$ is limited to $\mu<0.3~$GeV. At larger $\mu$, the resulting $s/T^3$ exceeds the characteristic value. This is an evidence that the two freeze-out conditions, $s/T^3$ and  $\langle E\rangle/\langle N\rangle$ can at least partly be explained by the gravitational deconfinement and the Hawking-Unruh radiation from charged black holes. We found that the entropy density $s$ remains independent on $\mu$ (or $Q$). As discussed in literature \cite{Tawfik:2005qn,Tawfik:2004ss}, $s$ describes among others the degrees of freedom of the system. The results in Fig. \ref{Fig2} would be interpreted that the charged black hole's degrees of freedom are not affected by the Hawking-Unruh radiation. Furthermore, no information loss accompanies such a radiation process. Indeed, from the area-proportionality of black hole entropy and the entanglement between the field's degrees of freedom inside and outside the horizon, the degrees of freedom of the field's ground state near the horizon  contribute to a large extent to the entropy and the area-proportionality is verified \cite{BHdof}.

A further systematic investigation for the analogy between the hadron production in high-energy collisions and the Hawking-Unruh radiation process is planned to be studied in the near future. It intends to study the thermodynamical properties as pressure, energy density, and trace anomaly from Hawking-Unruh radiation.


\end{document}